\documentclass[11pt,aps,prd,preprint,preprintnumbers,showpacs,superscriptaddress,floatfix]{revtex4-1}
\usepackage{graphicx}
\usepackage{epsfig}
\usepackage{dcolumn}
\usepackage{bm}
\usepackage{subfigure}
\usepackage{amsmath}
\usepackage{amssymb}
\usepackage{bbm}
\usepackage{multirow}
\usepackage{setspace}
\usepackage{xcolor}
\usepackage[colorlinks,
            linkbordercolor={0 0 0},
            pdfborder={0 0 0},
            linkcolor=blue,
            citecolor=blue,
            urlcolor=blue,
            breaklinks=true]{hyperref}

\def \tct{\textcolor{teal}}

\begin{document}
\title{Initial fluctuations and power spectrum of flow anisotropies
in relativistic heavy-ion collisions}
\author{Shreyansh S. Dave}
\email{shreyanshsd@imsc.res.in} 
\affiliation{The Institute of Mathematical Sciences, Chennai 600113, India}

\author{Saumia P.S.}
\email{saumia@theor.jinr.ru} 
\affiliation{Bogoliubov Laboratory of Theoretical Physics, JINR, 141980 Dubna, Russia}

\author {Ajit M. Srivastava}
\email{ajit@iopb.res.in}
\affiliation{Institute of Physics, Bhubaneswar-751005, India}
\begin{abstract}
Flow has emerged as a crucial probe for the properties of the 
thermalized medium produced in relativistic heavy-ion collisions. The 
evolution of initial state fluctuations leaves imprints on the power 
spectrum of flow coefficients. Therefore flow coefficients are a crucial 
probe of initial state fluctuations arising from the parton distributions 
of the colliding nuclei. This has a very strong correspondence with the 
physics of power spectrum of cosmic microwave background radiation 
(CMBR) anisotropies which directly probes initial inflationary fluctuations. 
Much work has been done to probe these interesting interconnections, in 
particular, in developing techniques for the measurements of higher 
flow coefficients. We present a short review of these developments. 
The effect of initial 
magnetic field on these features will also be reviewed. All this acquires 
special importance in view of upcoming electron-ion collider which will 
directly probe initial parton distribution of the colliding nucleus.

\end{abstract} 
\maketitle
\section{Introduction}
\label{intro}
Relativistic heavy-ion collision experiments (RHICE) have provided us with
a remarkable opportunity of investigating properties of strongly interacting
matter under extreme conditions of temperature and/or baryon density.
This complements our efforts to understand perturbative aspects of 
quantum chromodynamics (QCD), the theory of strong interactions, with
ultra high energy colliders, extending it in the regime where
non-perturbative aspects play crucial role.  Indeed, entire QCD
phase-diagram is now subject of experimental investigation with
issues like phase transition, critical point, etc., being examined in the
light of experimental data as well as theoretical predictions using
non-perturbative techniques like lattice gauge theory, effective field 
theory etc. All this has allowed us to make significant progress
in our overall understanding of QCD. At the same time we are able to 
study, under experimentally controlled situations, those aspects of our
universe which are beyond direct reach of experiments. 

High baryon density
regime of QCD is being probed by the beam energy scan program of RHIC, and
will be the main focus of upcoming facilities FAIR and NICA. These directly
provide us with inputs for understanding the behavior of matter in important
astrophysical objects such as neutron stars, and possibly the behavior of
stars undergoing collapse to black holes during their last stages. Exotic
phases of QCD have been postulated in the QCD phase diagram at very high
baryon densities such as color flavor locked (CFL) phase, crystalline
superconductivity phase, 2SC phase etc. which could occur in the interiors
of such objects \cite{cfl}. Possibilities are 
being explored of detecting such phases
in heavy-ion collision experiments. Even a somewhat more conventional,
nucleonic superfluidity phase, which is believed to be crucial for 
understanding pulsar glitches, may become accessible in relatively low
energy heavy-ion collisions \cite{qcdsf}.

 A completely different regime in the QCD phase diagram provides a
 direct insight into the very early stages of our universe, when its 
 age was about few microseconds. This is the regime of  high 
 temperature and very low baryon density. Ultra-relativistic 
 collisions of heavy nuclei at RHIC and LHC have provided, and are
 continuing to provide, invaluable data which have made qualitative 
 changes  in our understanding of this extremely important regime of 
 QCD  phase diagram. This is the regime in which lattice QCD simulations 
 have provided extremely reliable calculations (compared to the high 
 baryon density regime). A constant dialogue between lattice predictions
and experimental observations have allowed reasonably reliable conclusions
to be reached, e.g. the formation of quark-gluon plasma 
(QGP) phase of QCD in these 
experiments and the quark-hadron transition temperature. It has shown
that quark-hadron transition in this regime of phase digram is a crossover 
transition. The correspondence with  the early universe phase certainly makes 
this regime very exciting.

 Probably the most important observation from the
 relativistic heavy-ion collision experiments is the measurement of
 so called {\it elliptic flow} \cite{v2}. There have been many signals
 proposed for the observation of the quark-gluon plasma (QGP) phase of QCD in
 these experiments. Starting with the $J/\psi$ suppression, to 
 strangeness enhancement, jet quenching, photons/dileptons are some
 of the important signals which have been thoroughly analyzed and
 compared with data with varying degrees of success in providing
 a clean signal for QGP formation. Certainly, all the signals 
 together, including  elliptic flow, have allowed us to be
 confident that indeed the QGP phase has been produced in these
 experiments. At the same time it seems fair to draw attention to
  elliptic flow (flow in general) in providing us with qualitatively
 new features of the thermalized medium produced. Two points can
 be made to support this claim, the first one being thermalization. All 
 other signals require quantitative details to distinguish
 between the effects of a thermalized medium from the effects of
 a dense medium which may be out of equilibrium. However, elliptic flow
 most directly probes the equilibrium behavior of the medium. As we
 will explain below, whatever be the anisotropies in the initial spatial 
 distribution of energy density, in an ultra-relativistic collision where
 initial transverse velocity is negligible, momentum anisotropies can only
 arise from development of anisotropic pressure gradients. Thus, a degree
 of equilibration is necessary so that well defined distribution of
 pressure can arise. Though there have been efforts to explain 
 the observed momentum anisotropies in terms of anisotropic
 diffusion through a dense medium, without assuming equilibrium, such
 efforts have not met much success in accounting for the wealth of
 data on elliptic flow.

  The second point because of which elliptic flow needs special mention
  is the qualitatively novel behavior of QGP it has revealed, way beyond
  any theoretical expectation. All other signals have only aspired to
  probe the standard picture of QGP as a thermalized gas of deconfined 
  quarks and gluons. Elliptic flow has directly probed a very important 
  transport coefficient, namely $\eta/s$, the shear viscosity to entropy
  density ratio. The experimental data is consistent with hydrodynamic
  simulations   only with very small values of $\eta/s$, very close to
  the lowest limit $1/4\pi$ \cite{adscft}. This is the smallest value 
  of all known
  liquids, making QGP in these experiments as the most ideal liquid ever
  produced. This was certainly totally unexpected. Indeed, it is even
  contrary to the original spirit of the hypothesis of QGP where one argued 
  for the existence of a weakly interacting deconfined gas of quarks
  and gluons at very high temperatures based on the asymptotic freedom 
  of QCD. Instead, what one is seeing is that at the temperatures produced in
  these experiments, QGP is far from being an ideal gas (which should have
  large mean free path, hence large shear viscosity), but is behaving
  like a strongly interacting/correlated system.

  It is then not surprising that elliptic flow, and flow in general has 
  taken, in some sense, a center stage in the investigation of QGP in
  RHICE. A very important realization in this regard was about the 
  importance of initial state fluctuations in energy density. Due to 
  random phase space
  distributions of nucleons (and partons within) inside colliding nuclei, 
  the resulting initial medium necessarily had inhomogeneities in the 
  transverse plane. It was well recognized that in calculations of elliptic
  flow $v_2$, as well as certain higher flow coefficients (namely $v_4$, 
  $v_6$,  and very occasionally $v_8$) in a non-central collision,
  there are  uncertainties arising from the error in 
  defining the axes of the event-plane due to these fluctuations. Many 
  investigations were carried out on these issues and techniques were 
  developed to take care of these effects. It was also recognized that
  these flow coefficients may have small non-zero values even in central 
  collisions due to these initial fluctuations. It is interesting that 
  despite this recognition of effects of fluctuations, no attention was 
  paid to the other flow harmonics. In particular, odd flow coefficients were 
  completely neglected.

 A very different view on these initial state fluctuations was initiated
by some of us in a series of papers \cite{cmbhic,vnrms}.  The 
QGP, produced in RHICE, has initial energy density fluctuations. Because of 
presence of inside-outside pressure gradient, it expands hydrodynamically, 
therefore cools down and reaches quark-hadron transition temperature, where 
QGP to hadron crossover transition occurs. These hadrons further evolve, and 
first chemically, then thermally freeze out, and finally reach the detectors 
 carrying certain momentum distribution in the transverse plane. This 
momentum distribution of hadrons carries imprints of the intial state 
fluctuations and the properties of medium. Indeed, in discussions of heavy-ion 
collisions, it is often mentioned in popular terms that attempts to learn 
about the phase of matter in the early stages from the observations of hadrons 
is similar to the attempts to understand the early stages of the 
universe from the observations of the cosmic microwave background radiation 
(CMBR). The surface of last scattering for CMBR is then similar to the 
freezeout surface in RHICE. The last scattering surface 
represents the time when protons and electrons `recombine' and the universe
becomes neutral enabling the photons to free stream through the universe.
In refs.\cite{cmbhic,vnrms}, such 
qualitative statements were 
extended to a deeper level of correspondence between flow fluctuations in RHICE 
and the CMBR fluctuations in the universe. Following the successes of the 
analysis of the CMBR anisotropy power spectrum in providing crucial information 
about initial inflationary density fluctuations, it was argued in these works 
that flow coefficients should be used as a probe for identifying initial state 
fluctuations in RHICE, thereby providing crucial information about initial 
nucleon/parton distributions. Thus, initial fluctuations should not only be 
considered as providing errors in calculating certain flow coefficients for 
non-central collisions, they should be the main focus of study as a source of 
information about initial system itself. From that point of view, central 
collisions became much more important, as a large peak at $v_2$ for non-central 
collisions becomes a distractor when the focus is only on the initial state 
fluctuations. For a central collision, all flow coefficients became important, 
including the odd flow coefficients.  With that, it was argued 
in \cite{cmbhic,vnrms}, that one should plot the power spectrum of all flow 
coefficients, with the entire plot providing crucial inputs on the initial 
state fluctuations, as well as their evolution.

 It is now generally recognized that a large number of flow coefficients
need to be studied which not only contain effects of initial fluctuations, 
but also important correlations arising from hydro evolution. 
The subject 
of this short review is to provide developments in this area of
study of flow coefficients with special focus on their power spectrum. We 
will begin in Sect.2, with a brief recollection of the 
importance of elliptic flow and the effects of initial 
fluctuations on the determination of specific even flow coefficients in 
terms of resulting uncertainties  in the determination of the event plane.
Sect.~\ref{3:1} presents the new perspective on the initial state 
fluctuations as proposed in refs.\cite{cmbhic,vnrms}
emphasizing the importance of power spectrum of flow coefficients. Here we will 
draw correspondence with the power spectrum of CMBR anisotropies and discuss 
possibilities of similar features, such as CMBR acoustic peaks in the flow 
power spectrum.  Here we discuss results from several investigations 
where general study of effects of initial state fluctuations on flow 
coefficients has been carried out. In Sect.4, we discuss some studies 
where correspondence with CMBR studies has been further explored.
In Sect.~\ref{5:1} we discuss ways to isolate the effects of  initial state 
fluctuations from the effects of hydrodynamical evolution. For this we present 
results of magnetohydrodynamical evolution which show qualitative patterns 
on the power spectrum of flow coefficients in the presence of very strong 
magnetic fields. As the magnetic field is expected to be very strong only for 
very early stages (subsequently slowly decaying in time with medium effects 
included), such qualitative features of flow power spectrum 
can provide unique probe of the magnitude of initial state fluctuations 
which will be the subject of main focus for the upcoming electron-ion collider. 
We also discuss such qualitative patterns arising from any
superfluid phase of QCD which could be produced in relatively low energy 
collisions. In Sect.~\ref{6:1} we will conclude with discussion 
on new directions. 

\section{The elliptic flow}
\label{2:1}
 
Elliptic flow has yielded the very useful and surprising information that 
the matter formed at RHIC behaves like an ideal liquid. In a simple picture, 
for non-central collisions, the interaction region is not circular in 
the transverse plane ($xy$-plane as shown in Fig.1), but rather has an 
elliptical shape. Once thermalization is achieved, the formed fluid has a 
thermal pressure, which varies in space with maximum value at the center 
of the system and zero outside in vacuum. Clearly the pressure gradient, in 
the transverse plane, will be larger along the semi-minor axis of the 
ellipse (taken to be the $x$-axis in Fig.1). This forces the plasma to 
undergo hydrodynamic expansion at a faster rate in that direction compared 
to the semi-major axis (the $y$-axis in Fig.1). Thus particles reach the 
detectors with larger momenta along the $x$-axis than the $y$-axis. In 
other words, the spatial anisotropy gets transferred into momentum 
anisotropy due to hydrodynamical flow. 
\begin{figure}
\begin{center}
\resizebox{0.75\columnwidth}{!}{%
\includegraphics{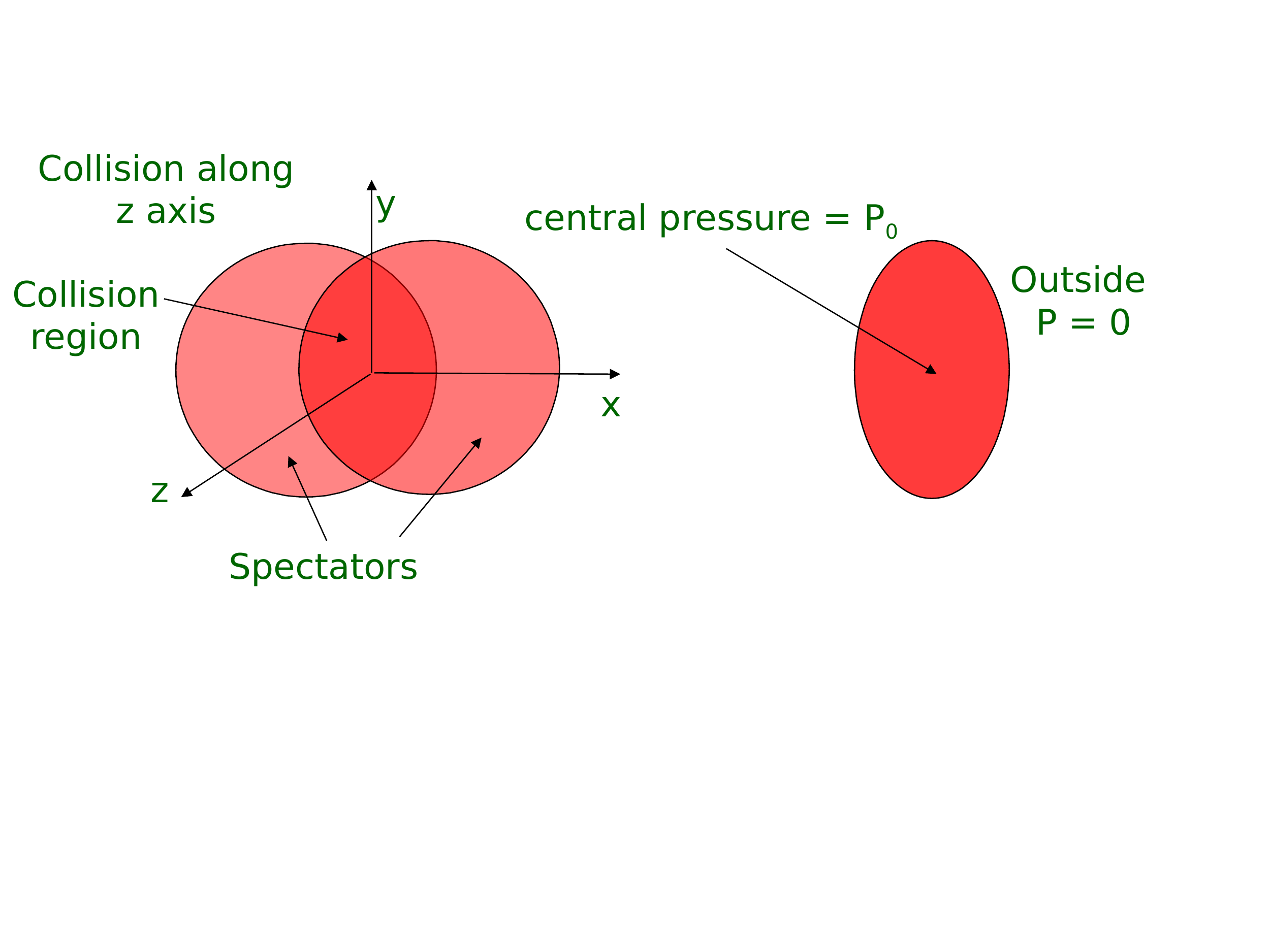} }

\caption{Non-central collision leads to anisotropic interaction region in the transverse plane ($xy$-plane). 
With thermalization one expects formation of an elliptical region of QGP in the transverse plane as shown on 
the right of Fig.1. With anisotropic shape, and no initial transverse expansion, anisotropic pressure gradient 
implies that  buildup of plasma flow will be larger in the $x$-direction than in the $y$-direction, leading to 
generation of elliptic flow.}
\label{fig1}
\end{center}
\end{figure}

Clearly the generation of elliptic flow depends crucially on the equation of 
state relating pressure to the energy density and transport coefficients,
e.g. shear viscosity to entropy ratio $\eta/s$. Thus, the observed momentum 
anisotropy of the particle distribution can be used with hydrodynamical
simulations to extract useful information about hydrodynamic flow at very 
early stages thereby directly probing $\eta/s$ and the equation of state of 
the QGP (usually taken from lattice results).  It is important to note that 
elliptic flow gives  probably the most direct estimate of the thermalization 
time. If thermalization is delayed by a certain time, the elliptic flow would 
have to build on a reduced spatial deformation and would come out smaller.
The observations put an upper limit of about 1 fm on the 
thermalization time for ultra-relativistic collisions at RHIC and LHC energies.
The experimental data seems to be in very good agreement with the 
prediction of almost ideal hydrodynamics pointing to a very 
low $\eta/s$ of the QGP produced. This shows that the QGP does not behave as 
a weakly interacting quark-gluon gas as predicted by perturbation theory, 
rather it behaves as a strongly interacting/correlated
liquid. This is termed as Strongly Coupled QGP (sQGP), with a strong 
non-perturbative interactions/correlations.

Anisotropy in the transverse momentum distribution is captured by the
  flow coefficients which are the Fourier coefficients of the azimuthal
  momentum distribution of particles. We consider the Fourier series of
  the azimuthal distribution of fractional transverse momentum
  distribution \cite{Voloshin:1994mz} 

  \begin{equation}
  {1 \over {\bar {p_T}}}{dp_T(\phi) \over d\phi} =
\sum_{n=0}^{\infty} \bigg(a_n \cos(n\phi) + b_n \sin(n\phi)\bigg),
  \end{equation}

Here, $p_T(\phi)$ is the net transverse momentum in the angular bin at
azimuthal angle $\phi$ and ${\bar {p_T}}$ is the angular average of the
transverse momentum.  The flow coefficients $v_n$ are appropriate event 
averaged values of $a_n$ and $b_n$. This definition of flow coefficients can 
be directly used for particle distributions as well as for the fluid
momentum distributions in hydrodynamic simulations. We write the complete 
expansion here in the anticipation of the presence of fluctuations. In the 
absence of fluctuations, there is a reflection symmetry with respect to 
the reaction plane with which only the cosine terms survive. 
Generalization to transverse momentum $p_T$ and rapidity $y$  dependent 
flow coefficients $v_n(p_T,y)$ can be written straightforwardly in terms 
of differential distributions.

Even though we use the above definition for elliptic flow, very often
the flow coefficients are defined as the Fourier coefficients of the
azimuthal distribution of the final particle number. These two quantities 
have a straightforward correspondence since larger momentum in a bin 
in a fluid means a larger number of particles flowing into that bin.
There are several methods of measuring the elliptic flow which is the
second Fourier coefficient in the definition above. 
One method is to estimate the event plane, and then correlate the outgoing particles 
to this plane (for detail see, \cite{ep1,ep2}). One could also use a two particle 
correlation method to calculate the elliptic flow \cite{Wang:1991qh}.
 These 
two methods are equivalent even though the latter does not need the 
determination of event plane.  But it has been shown that both these 
methods have limitations due to event-by-event fluctuations as well as 
presence of non-flow correlations arising from resonance decays, jet 
fragmentation etc. The picture of a smooth elliptical QGP region for a 
non-central collisions (as in Fig.1)
leading to elliptic flow is too simplistic.  It was 
well known that initial state fluctuations are always present for any 
centrality. Due to these fluctuations, initial energy density distribution in 
the QGP region is non-homogeneous, e.g. as shown in Fig.2 for a central 
collision. Multiparticle cumulant expansions have been proposed to take care 
of these as well as detector effects \cite{Borghini:2001vi}.
It has been shown that there are improved methods involving multiparticle 
correlations like four particle cumulants and using the event plane 
determined from directed flow in a zero degree calorimeter using three 
particle correlations of the spectators. These are insensitive to non-flow 
correlations as well as initial eccentricity fluctuations and hence measure 
elliptic flow effectively \cite{Bhalerao:2006tp,Ollitrault:2009ie}.


\begin{figure}
\begin{center}
\resizebox{0.25\columnwidth}{!}{%
\includegraphics{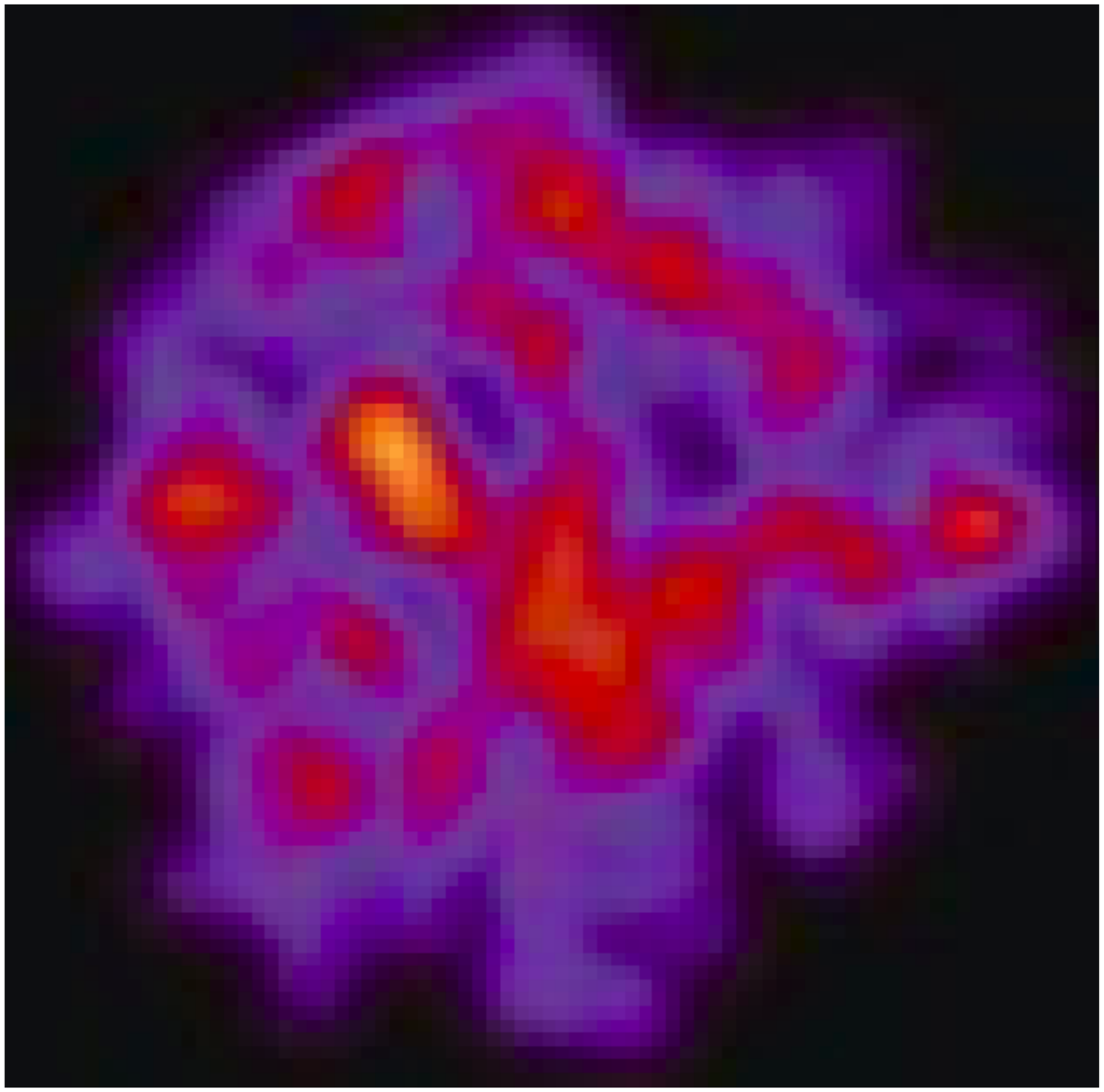} }
\caption{With initial state fluctuations necessarily present, the resulting
QGP region is not homogeneous, rather it is lumpy as shown here for a 
	central collision \cite{blaizot}}
\label{fig2}
\end{center}
\end{figure}

\section{Correspondence with CMBR and the power spectrum of flow anisotropies}
\label{3:1}

 As we mentioned above, initial state fluctuations were initially discussed 
primarily in the context of  determination of the event plane for elliptic 
flow calculations. Higher harmonics like $v_4$ and $v_6$ were seen
as induced from $v_2$ and the eccentricity fluctuations as a higher order 
effect. However, no other flow coefficients, in particular, no 
odd harmonics were discussed except a very early mention of the possibility of 
$v_3$ as well as $v_4$ due to initial deformation of the colliding nuclei 
rather than initial fluctuations \cite{Voloshin:1994mz}.  The 
main reason for this was that the focus primary remained on non-central 
collisions to get elliptic flow which gave information about very important 
properties of QGP phase such as equation of state, viscosity etc.

 This view towards initial state fluctuations, as nuisance in getting the 
values of flow coefficients was reversed in a series of papers by some of 
us \cite{cmbhic,vnrms} where these initial state fluctuations were made the center
of attention. In these works it was argued that initial state fluctuations 
are extremely important, originating from initial conditions, namely parton 
distributions inside  the colliding nuclei. Thus it was argued, in particular, 
that the central collisions are very important. Non-central collisions 
retain their importance in getting strong signal for elliptic flow which
probed equation of state, shear viscosity to entropy density ratio etc. However,
when one wants to learn about the initial state fluctuations,  it is better 
to focus on central collisions since the very large elliptic flow in 
non-central collisions tends to mask the effects of initial state fluctuations. 
 
 Fig.2 shows the typical initial energy density distribution for a
central collision at the thermalization stage.
As one can see, inhomogeneities of all scales are present, even 
in central collisions. With such lumpy initial energy density distribution,
hydrodynamical evolution will be expected to lead to all flow coefficients 
becoming non-zero in general. Thus, all Fourier coefficients  $v_n$
should be of interest, including odd harmonics. 
The fact that the fluctuations and anisotropies in the final particle 
momentum distribution is directly related to fluctuations in the 
initial energy density distribution was further studied by various works 
\cite{Mocsy:2010um,Sorensen:2010zq,Alver:2010gr,Qin:2010pf}. There have been 
many discussions about various ways of  appropriately quantifying the 
initial state fluctuations, e.g. see 
\cite{Alver:2010gr,Qin:2010pf,raju}, so that the higher harmonics can be 
explained as a response to them. The higher flow harmonics are experimentally
measured and their correspondence with different initial condition models were 
studied in \cite{ALICE:2011ab,Adare:2011tg,Aad:2013xma,Lacey:2010hw}. Various 
flow observables including the ratios of different harmonics are shown to be 
largely determined by the initial state and hence helpful in studying the early
stages \cite{Bhalerao:2011yg}. It was also shown using viscous hydrodynamic 
simulations and other models that the different modes couple non-linearly 
during the evolution \cite{Qian:2016fpi,Bravina:2013xla,Sirunyan:2019izh}.
 
It is worth pointing out that this shift in focus to initial state fluctuations 
in ref.\cite{cmbhic,vnrms} was motivated from the realization of deep 
similarities between the physics of 
flow anisotropies in heavy-ion collisions and CMBR anisotropies in the universe. 
In this section we will explain these motivations and develop this very 
intriguing correspondence in detail. To the skeptic reader we mention that one 
main difference between the two system is the absence of gravity for
heavy-ion physics. It will be easily seen below that it only affects overall
scale of the resulting distribution of flow coefficients (the power spectrum
of flow coefficients), without having any important effect on its shape.
Another important difference is the presence of strong interactions in RHICE
compared to the CMBR case where the physics at last scattering surface is 
governed by only electromagnetic interactions.
As a result equation of state of the matter is different in the two cases, but again, this 
is only expected to affect the quantitative features of the shape of the power 
spectrum.

 As we mentioned, it has always been appreciated that the surface of
last scattering of CMBR is in many ways like the freezeout surface for heavy-ion
collisions. This is in the sense that for the former case, one can learn 
about the early universe from the CMBR photons from the surface of last 
scattering.  In the same way, for heavy-ion collisions, one only gets
hadrons from the freezeout surface. It is these hadrons which have to be 
analyzed to learn about the QGP system. The main ingredient in the new
approach to flow coefficients relates to the fact that CMBR fluctuations 
originate from inflationary fluctuations during initial stages of the
universe. With CMBR power spectrum, one is able
to learn about these initial inflationary fluctuations. In fact, the
later stages of the universe (post-inflation) simply evolve these fluctuations.
This evolution has to be understood so that one can isolate the primordial 
inflationary fluctuations. In the same way, for heavy-ion collisions also, a power
spectrum of flow coefficients, should be used to probe directly the initial
state fluctuations, with proper account of medium evolution effects.  

 This change of perspective naturally invites the use of techniques of CMBR 
analysis for heavy-ion case.  For CMBR, the
temperature anisotropies are analyzed using spherical harmonics,
as appropriate for the surface of 2-sphere (the CMBR sky) \cite{cmbr}.

\begin{equation}
{\Delta T \over T}(\theta,\phi) = a_{lm} Y_{lm}(\theta,\phi),
\end{equation}

where, $T$ is the average CMBR temperature and $\Delta T$ is the fluctuation 
in the temperature from its average value.
The coefficients of the expansion $a_{lm}$, corresponding to
the spherical harmonic $Y_{lm}$, are degenerate in the argument $m$. 
When averaged over different values of $m$,  these vanish
due to isotropy of the universe.  The variance of $a_{lm}$
denoted by $C_l$ (with suitable normalizations) is plotted with respect
to $l$ leading to the celebrated 
power spectrum of CMBR anisotropies \cite{cmbr}.

\begin{equation}
<a_{lm}> = 0 ~~, C_l \sim <|a_{lm}|^2>.
\end{equation}

The same technique was applied in \cite{cmbhic,vnrms} for 
analyzing particle momentum anisotropies, using 
lab fixed frame, in RHICE to probe the flow anisotropies.
For RHICE, focusing on central rapidity region, one analyzes
momentum anisotropies on a circle, requiring the use of the
Fourier coefficients $v_n$.  These should be distinguished
 from the conventional flow coefficients $v_n$ which are defined 
with respect to the  event plane. However, our purpose here is to develop
a probe of initial state fluctuations and these $v_n$ defined here serve
this purpose. For relation between these flow coefficients and the
conventional ones, see ref.\cite{vnrms}. With a fixed lab frame, the event
average values of these $v_n$s will all be zero due to rotational 
symmetry. We then use the variance of $v_n$, i.e. 
$v_n^{rms}$ in analogy with $C_l$ for CMBR. 

\begin{equation}
<v_n> = 0, ~~~ v_n^{rms} = \sqrt{<v_n^2>}
\end{equation}
where $<v_n^2>=<a_n^2>+<b_n^2>$;  
 $<..>$ denotes event average of the quantity, and $a_n$ and $b_n$ are
the Fourier coefficients in Eq.(1).  We point out here that a similar
  definition of power spectrum of flow coefficients was earlier proposed
  though it was in the context separating flow and 
  non-flow effects~\cite{Trainor:2007fu}.
  In view of the correspondence with the CMBR power spectrum, the flow
  coefficients were defined in ref.[8] using the azimuthal distribution
  of $\Delta p_T(\phi)/({\bar {p_T}}\Delta \phi)$ where $\Delta p_T(\phi)
  = p_T(\phi) - {\bar {p_T}}$ with  ${\bar {p_T}}$ being  the angular
average of the transverse momentum $p_T$.

A plot of $v_n^{rms}$ vs. $n$ for a large range of $n$ will provide the 
power spectrum of flow coefficients for relativistic heavy-ion collisions.
The detailed structure of this plot for central collisions should reveal 
information about initial state fluctuations as well as their hydrodynamical 
evolution. 

Several works later adapted CMBR analysis techniques more elaborately
for the study of flow fluctuations using spherical harmonic expansion of 
the momentum distribution of particles including the pseudorapidity $\eta$
\cite{Naselsky:2012nw,Sarwar:2015mma,Llanes-Estrada:2016pso,Sarwar:2017iax,Machado:2018xvi,Machado:2019iuu}. They discuss the
relation between the full angular power spectrum and flow coefficients. 
References \cite{Naselsky:2012nw} and \cite{Machado:2018xvi} also show the 
Molleweid projection of the momentum distribution similar to the WMAP and 
COBE maps of the cosmic microwave background radiation and propose that 
these maps can be used to study the non-flow fluctuations after subtracting 
out the collective effects. Using hydro simulations, maps 
of fluctuations of energy density and temperature in small phase space bins 
have been produced similar to CMBR maps \cite{tapan}.

With this important lesson from CMBR analysis tools for RHICE, the next
step is to ask what important features of this power spectrum can be expected, 
and if at all there can be any similarities with the shape of CMBR power
spectrum which is shown in Fig.3. We will summarise the relevant
parts of the physics of CMBR power spectrum below.

\begin{figure}
\begin{center}
\resizebox{0.6\columnwidth}{!}{%
\includegraphics{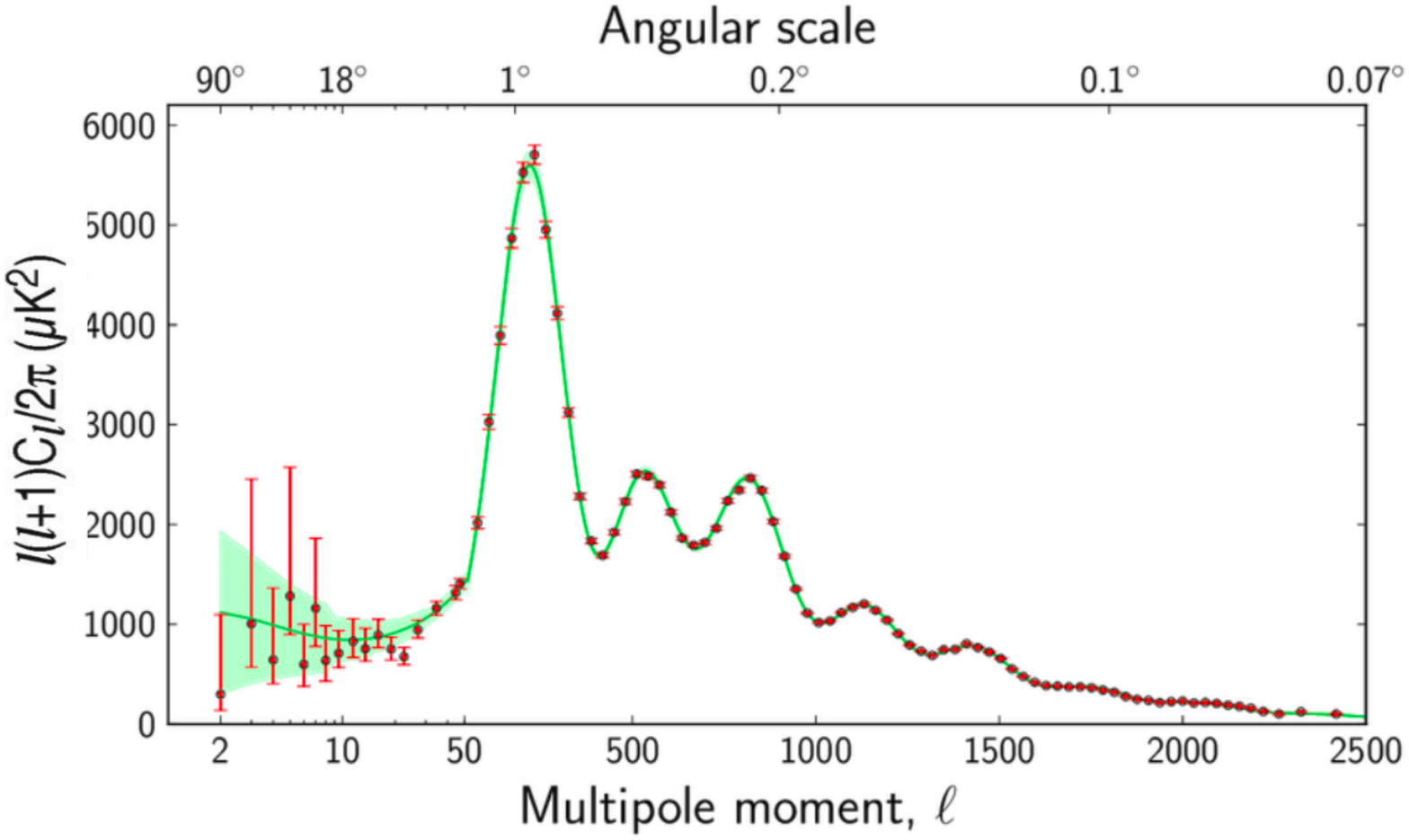} }
\caption{CMBR power spectrum}
\label{fig3}
\end{center}
\end{figure}

\subsection{The essential features of the CMBR power spectrum}

Fig.3 shows the CMBR power spectrum where $l$ along the X-axis is related to 
the different wavelenghth modes of the temperature fluctuations on 
the CMBR sky. $l$ could also be understood as $l \sim \pi/\theta$ where $\theta$ is 
the angular separation on the 
CMBR sky. The Y-axis corresponds to the power in each mode at the time of 
last scattering. The temperature fluctuations on the other hand, are directly related
to the density perturbations on the last scattering surface. These density perturbations     
have their origins in the early inflationary stage of the universe. During inflation,
the universe expanded exponentially, stretching out the quantum fluctuations in 
the inflaton field to superhorizon scales.  As inflation ends,
reheating occurs converting these fluctuations of  into matter density
perturbations. These perturbations evolve after inflation and  
 eventually
 lead to structure formation after matter domination stage.

One of the important features of these density perturbations is their superhorizon
length scales. In simple terms, for the universe the horizon size = speed of light 
$c$ $\times$ age of the universe $t$. The perturbations
stretched out by the inflation start re-entering the horizon after the inflation
owing to the faster expansion of the horizon scale ($\sim t$) as compared to the expansion
of the universe ($\sim t^{1/2}$ in radiation dominated era or $\sim t^{2/3}$ 
in matter dominated era ). As they enter the horizon,
the perturbations start oscillating. In the matter dominated era, the perturbations
'grow' due to gravitational effects and oscillate.
But the scales larger than the horizon size at recombination  stay as superhorizon 
modes in the CMBR temperature fluctuations on the last scattering surface. 
In Fig.3, $l \sim 200$ represents 
the horizon at last scattering. The lower $l$ modes represent the ones that are superhorizon. 
 These are the scales unaltered by any causal physics before the 
recombination stage in the universe except for damping effects due to photon diffusion 
and remain as laid down by the inflationary 
fluctuations.  Superhorizon fluctuations for universe do not 
oscillate (these are frozen). More importantly, they do not grow. That is, they are 
suppressed compared to the fluctuations which enter the horizon and grow by
gravitational collapse.

Another important feature of the inflationary density perturbations is their coherence.
As the fluctuations are stretched out of the horizon by inflation, they freeze out
dynamically. This means that any mode which is frozen out remains as an ocillation
in space, but not in time.
 As a result, when they re-enter, they enter with zero oscillation velocity  (in phase)
and hence modes of same wavelength start evolving coherently everywhere leading to coherent
acoustic oscillations. 

The location of the first peak in the power spectrum represents the largest mode 
which has grown to its maximum
at recombination and hence also tells us about the corresponding horizon scale.
$l>200$ represents the modes at the various stages of acoustic oscillations
at the time of last scattering. The positions and the heights of the 
different peaks in the power spectrum carry important information about the
contents of the universe until the time of last scattering.

\subsection{Power spectrum of flow anisotropies}	 

 We will focus on two main features of the power spectrum in Fig.3, namely
the suppression of superhorizon modes and the acoustic oscillations. We will argue
that similar physics is present for RHICE and hence these two features
should be present in the power spectrum of flow coefficients as well. First 
we discuss that probably the most important concept for the universe, that of 
a causal horizon, very naturally applies to the case of relativistic heavy-ion 
collisions.

We noted above (Fig.2) that initial state fluctuations of different length 
scales are present in relativistic heavy-ion collisions even for central 
collisions.  The process of  equilibration 
will lead to some level of smoothening. 
However, thermalization happens in a very short time scale. All estimates of
the thermalization time $\tau_0$ indicate very small values (as short as
a tenth of fm for LHC energies). Hydrodynamical simulations can
accommodate observed value of elliptic flow only with $\tau_0 < 1$ fm.
No homogenization can be expected to occur beyond length scales larger than 
$c \tau_0$ at this thermalization stage. This provides a natural concept of 
causal Horizon. The interaction region resulting from the collision of the two
highly Lorentz contracted nuclei is born at time $\tau = 0$ (definition of
origin of time for full overlap of nuclei). It takes a time $\tau = \tau_0$
for this system to thermalize leading to a locally equilibrated system
for which hydrodynamics becomes applicable. Relativistic hydrodynamics 
equations cannot lead to physical effects (of pressure differences etc.)
being communicated beyond the causal distance $c \tau_0$. More precisely for
the hydrodynamics, the limiting causal distance is the sound horizon
$c_s \tau_0$ where $c_s$ is the sound velocity (= 1/$\sqrt{3}$ for relativistic 
ideal plasma). Thus, inhomogeneities, especially anisotropies with 
wavelengths larger than this causal scale (horizon size) should be necessarily 
present at the thermalization stage when the hydrodynamic description is 
expected to become applicable.  With the nucleon size being about 1.6 fm, the 
equilibrated matter will necessarily have density inhomogeneities with 
superhorizon wavelengths at the equilibration stage. As time increases, the 
horizon size increases with time and larger  wavelength fluctuations become 
sub-horizon. The consequences of the presence of a sound horizon in the plasma
in different higher harmonics was also discussed later in~\cite{shuryak} 
where they also looked at the
effect of viscosity on the dissipation of different scales. 

 We will now discuss coherence and acoustic oscillations in case of RHICE. Coherence of 
inflationary density fluctuations essentially results from the fact that the 
fluctuations initially are stretched to superhorizon sizes and are 
subsequently frozen out dynamically. In the context of heavy-ion collisions, 
this freezing out is similar to absence of  initial transverse expansion 
velocity for QGP. Initially, fluctuations are only in spatial distribution of 
energy density, they become dynamical, converting to momentum anisotropies 
through hydrodynamical evolution. For all fluctuations of certain size 
$\lambda$, it happens ONLY after a certain time when horizon equals 
$\lambda/2$. Until then the fluctuations are almost frozen.  Thus coherence 
(meaning phase locking \cite{cmbr}) will be expected to 
hold for RHICE also. 

 Let us now discuss the oscillatory behavior for the fluctuations. We simply 
note that small perturbations in a fluid will always propagate as acoustic 
waves, hence oscillations are naturally present. It may seem surprising since 
typically, in the context of universe the oscillations are discussed in the
photon coupled baryonic system in the gravitational potential well of dark
matter. This is indeed the main difference for RHICE from the universe, 
the absence of gravity for RHICE. However, in the universe, the only role of 
attractive gravity  is to compress (collapse) the initial overdensities of 
cosmic fluid.  Acoustic oscillations happen on top of these collapsed 
fluctuations simply because the cosmic fluid is also governed by relativistic
hydrodynamical equations (in expanding universe). Similar relativistic hydro 
equations govern fluid evolution for RHICE also (with Bjorken longitudinal 
expansion in the early stages). Thus, for RHICE one will get harmonic 
oscillations (for a given mode) of plasma, while for the Universe one gets 
oscillations of a forced oscillator  (gravity acting as extra force) for the 
cosmic fluid. It can then be concluded that for RHICE also, there should be 
acoustic oscillations, which are coherent,  just as for CMBR. It is
important to realize that oscillations occur only for sub-horizon 
fluctuations. Only such fluctuations appear as perturbations in a background
which can propagate as a sound wave, for superhorizon fluctuations there
is not enough time for pressure gradients to lead to oscillatory behavior.

The smaller the length scale of the fluctuation, the earlier it will enter 
the horizon and start oscillating till the freezeout occurs. Hence the 
shortest scales will be most affected by any dissipative factors present in 
the system.  In the absence of any damping, a 
plot of $v_n^{rms}$ vs. $n$ should 
have acoustic oscillation peaks similar to the CMBR power spectrum 
with the value of $v_n^{rms}$ representing the stage of oscillation of 
the corresponding mode $n$ at freezeout. The peak structure of $v_n^{rms}$ 
vs. $n$ plot shows which mode has undergone dominant oscillations at 
the freezeout stage of the system. With time, various modes oscillate, 
depending on  dissipation present in the system.  Thus the higher harmonics 
will provide information on the dissipative properties of the medium.

 We now come to the second important feature of CMBR power spectrum: the 
behavior of modes which remain superhorizon at the surface of last scattering.  
 We have seen above that these modes are suppressed in CMB. For heavy-ion collisions, behavior of such 
superhorizon fluctuations will be extremely important as these will carry
information about  long range correlations in the initial state. These are
large wavelength modes corresponding  to low values of $n$ in the plot of
$v_n^{rms}$.  We now argue that for RHICE as well, there is a similar (though 
not the same, due to absence of gravity here) importance of horizon entering of modes.
 
One can argue \cite{cmbhic,vnrms} that flow anisotropies for superhorizon 
fluctuations in heavy-ion collisions should be suppressed by a factor of order 
$H^s_{fr}/(\lambda/2)$ where $H^s_{fr}$ is  the sound horizon at the  
freezeout time $\tau_{fr}$ ($\sim$ 5-10 fm for typical Pb-Pb collision), and
$\lambda$ is the wavelength of fluctuation.  This is because in heavy-ion 
collisions, spatial variations of density are not directly detected.
This is in contrast to the Universe where the spatial density fluctuations 
are directly detected in terms of angular variations of CMBR temperature.
For heavy-ion collisions, spatial fluctuation of a given scale (i.e. a 
definite mode) has to convert to fluid momentum anisotropy of the 
corresponding angular scale. This will get imprinted on the final hadrons 
and will be experimentally measured.  This conversion of spatial anisotropy 
to momentum anisotropy (via pressure gradients) is not effective for 
superhorizon modes. Thus, superhorizon modes will be suppressed in heavy-ion 
collisions. It will be very important to understand suppression of low n 
harmonics as these will contain the information about freezeout horizon size
as well as about long correlations at the initial stage. 

 We will see below that  results from relativistic hydrodynamical simulations 
support this suppression of long wavelength (low $n$) modes in the power 
spectrum of flow coefficients \cite{hydro}.

\section{Study of higher flow coefficients}
\label{4:1}

 Extensive experimental effort has gone in determination of higher flow
coefficients. The techniques typically require many-particle correlation
methods as discussed above in Sect.3. The first ever plot for a large values
of flow coefficients was presented by Sorensen in \cite{Sorensen:2008dm} 
and it was claimed that the plot shows suppression of superhorizon modes as 
predicted in \cite{cmbhic,vnrms}. Many experimental results have appeared 
since then. We show two sets of plots from ATLAS in Fig.4 and ALICE in Fig.5.

\begin{figure}
\begin{center}
\resizebox{0.7\columnwidth}{!}{%
\includegraphics{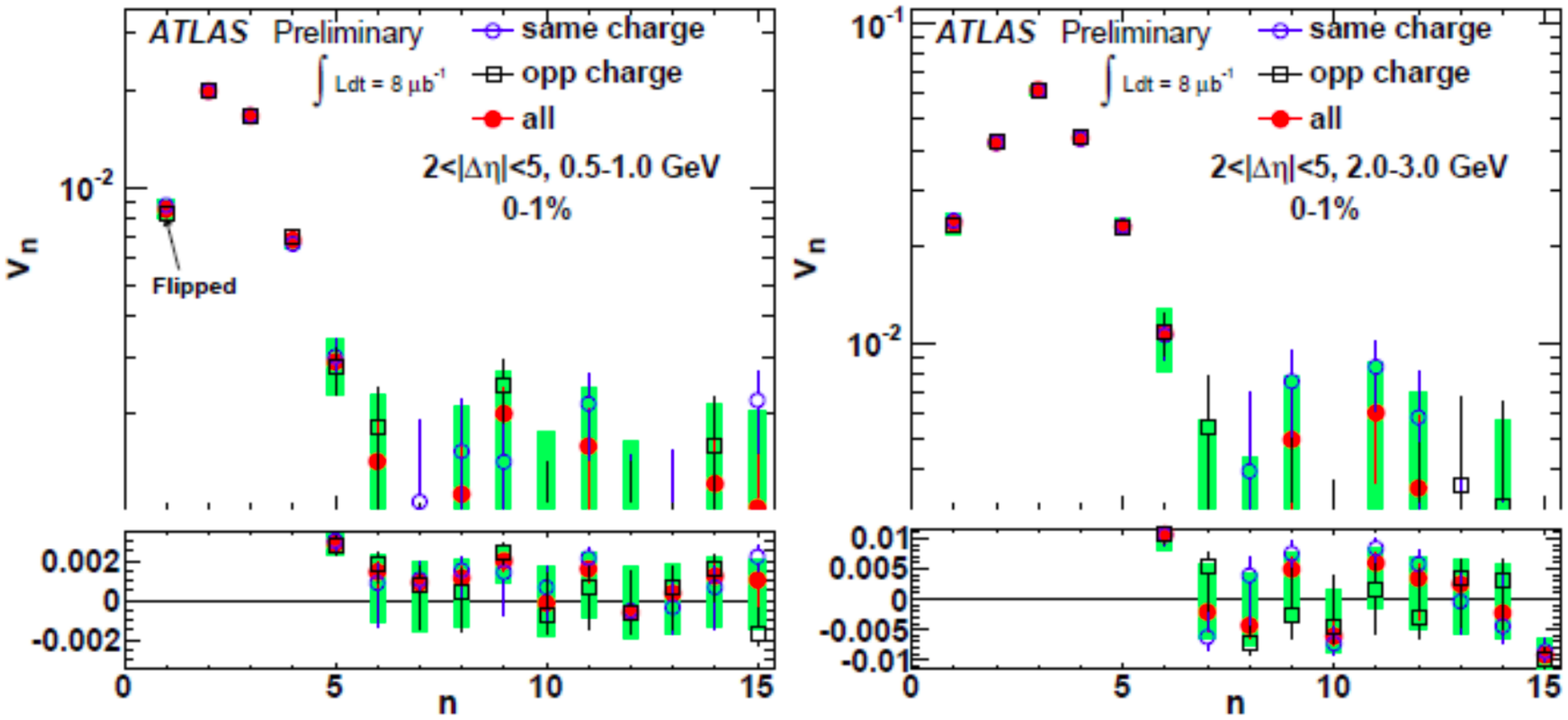} }
\caption{ Plot of a large range of flow coefficients for Pb-Pb collision
at 2.76 TeV at LHC. ATLAS Preliminary, 2011. Note, errors in $v_n$ become 
very large for $n \ge 6$. (Figure from arXiv: 1107.1468). We point out
the anomalous suppression of $v_2$ in the right plot. Such a suppression
of low $n$ modes was predicted in ref.\cite{cmbhic,vnrms} (termed as the 
{\it superhorizon suppression}).  Suppression of $v_1$ is also important to 
note. Although, being the directed flow,  it does not evolve  as a sound 
mode governed by hydrodynamical evolution. It retains its initial value 
determined by initial state fluctuations, hence suppressed compared to 
higher modes which can grow due to hydrodynamical evolution.}
\label{fig4}
\end{center}
\end{figure}

\begin{figure}
\begin{center}
\resizebox{0.8\columnwidth}{!}{%
\includegraphics{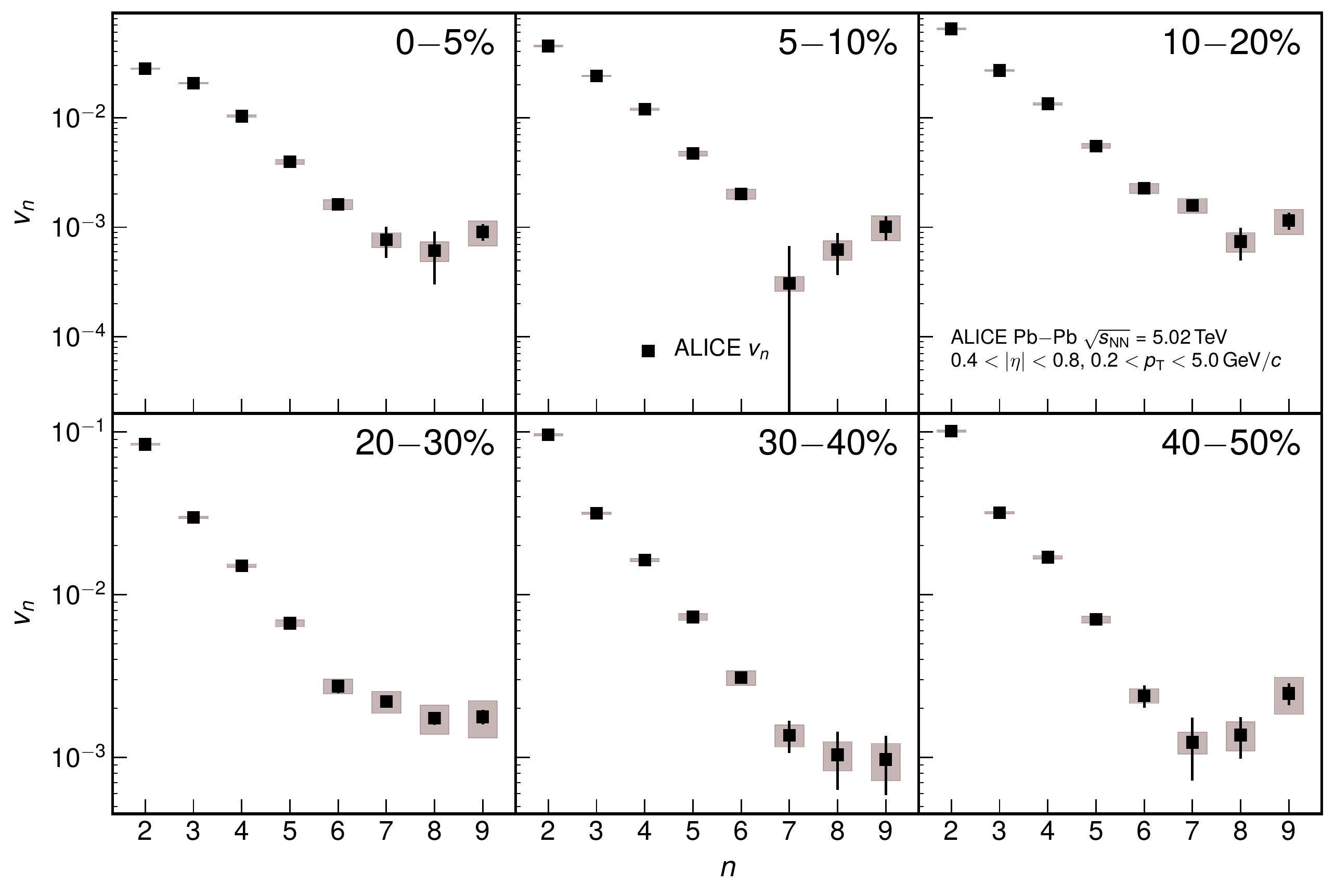} }
\caption{ Recent data set from ALICE (2020), showing plots of $v_n$ for
$n = 2 - 9$ for different centralities for  Pb-Pb collision at 
$\sqrt{s_{NN}}$ = 5.02 TeV. (Figure from arXiv: 2002.00633). Note that
the anomalous suppression of $v_2$ in the right plot in Fig.4 
(the {\it superhorizon suppression} \cite{cmbhic,vnrms}), is not seen in 
the plots in Fig.5. Also note that value of $v_1$ in Fig.4 also shows 
suppression,  while $v_1$ is not shown in Fig.5. (Note the comments about 
$v_1$ suppression in Fig.4). For the comparison of Fig.4 and Fig.5,
we point out  that, while Fig.5 shows the latest data which is more
refined, a direct comparison of Fig.4 and Fig.5 is difficult as the collision
energies and centrality coverage are different for the two sets. As discussed
in the text, physics of acoustic peaks can crucially depend on these. Most 
important feature for plots in Fig.5 is the rise of $v_n$ for $n > 7-8$. 
This rise will be entirely unexpected from general considerations of
flow coefficients, while this is exactly the behavior expected from the 
presence of acoustic peaks in the power spectrum (similar to the CMBR case) 
as predicted in refs.\cite{cmbhic,vnrms}).} 
\label{fig5}
\end{center}
\end{figure}

Fig.4 shows the plot of a large range of flow coefficients for Pb-Pb 
collision at 2.76 TeV at LHC. (ATLAS Preliminary, 2011, figure from arXiv: 
1107.1468, ref.\cite{atlas11}). Though it may be tempting to see some sort of 
acoustic peak like behavior in these plots for higher values of $n$, we note
that errors in $v_n$ are very large for $n \ge 6$ to reach any such conclusion. 
We thus focus on the plot for low values of $n$ for which errors are in better 
control. The suppression of $v_2$ in  the right plot is anomalous. 
This is exactly the behavior which was predicted in ref.
\cite{cmbhic,vnrms} where suppression of low $n$ modes was termed as the 
{\it superhorizon suppression}.  Suppression of $v_1$ is also important to 
note. Although, being the directed flow,  it does not evolve  as a sound 
mode governed by hydrodynamical evolution. It retains its initial value 
determined by initial state fluctuations, hence suppressed compared to 
higher modes which can grow due to hydrodynamical evolution.

Unfortunately, this suppression of $v_n$ for low
values of $n$ is not seen in plots of $v_n$ in Fig.5 which shows recent data 
set from ALICE, 2020, showing plots of $v_n$ for $n = 2 - 9$ for different 
centralities for  Pb-Pb collision at $\sqrt{s_{NN}}$ = 5.02 TeV. (Figure from 
arXiv: 2002.00633, ref.\cite{alice20}). While the data set of Fig.5 is 
the latest data and is more refined, it may be noted that a direct comparison 
of Fig.4 and Fig.5 is difficult as the collision energies and centrality 
coverage are different for the two sets. Further, note that $v_1$ is not 
shown in Fig.5, while it is given in plots in Fig.4. For the left plot in 
Fig.4, it is the value of $v_1$ which shows the suppression.  The right plot 
in Fig.4 shows $v_2$ also to be suppressed (such a suppression of low $n$ 
modes during hydrodynamical evolution was termed as the {\it superhorizon 
suppression} in ref.\cite{cmbhic,vnrms}), while this is not seen in the plots 
in Fig.5. It is important to emphasize here that the presence of first peak 
and its location, in the power spectrum, crucially depends on the freezeout 
time (size of acoustic horizon) and the wavelength of fluctuation being 
considered.  Clearly, these will depend on the collision energy, as well as on 
centrality, along with the nature of initial state fluctuations.

At the same time, Fig.5 shows something very important, which could not be 
clearly seen in Fig.4 (due to large errors for $v_n$ for large values of $n$ 
in Fig.4). Note the rise of $v_n$ for $n > 7-8$ in Fig.5 for all
centralities except two cases (20-30\% and 30-40\%). Such a behavior is
entirely unexpected from general considerations of flow coefficients. But,
this is exactly the 
behavior expected from the presence of acoustic peaks in the power spectrum 
(similar to the CMBR case) as predicted in refs.\cite{cmbhic,vnrms}). Errors
for $v_n$ for these larger values of $n$ are in good control. This
is the first clear indication of the presence of acoustic peaks in heavy-ion
collisions.

We would like to conclude from Fig.4 and Fig.5 that the data shows hints 
of non-trivial physics in the plots of $v_n$ for a large values of $n$.
 Some of this has qualitative behavior of acoustic peaks in the power
 spectrum, and possibly even the presence of superhorizon suppression.
 Both these features need focused attention, as these can open a new
 direction to probe the early stages of the collision, in particular
 the spectrum of fluctuations and the medium  properties. We now present  
 results of hydrodynamics simulations \cite{hydro} which show these two 
 important features in the power spectrum of $v_n$, namely the acoustic peaks 
 and suppression of power for low $n$ mode (the superhorizon suppression).

 We take QGP system in the ideal hydrodynamics limit with the
energy momentum tensor of perfect fluid form,

\begin{equation}
 T^{\mu \nu} = (\epsilon + P) u^\mu u^\nu + P \eta^{\mu \nu}
\end{equation}

where $\eta^{\mu \nu}=diag(-1,1,1,1)$ is the Minkowski space-time metric,
$\epsilon$ is the energy density and $P$ is the pressure. 
$u^\mu$ is the 4-velocity of the fluid. 

  Conservation of the energy-momentum tensor gives the
equations for ideal relativistic hydrodynamics,

\begin{equation}
\partial_\mu T^{\mu \nu} = 0
\end{equation}

We take ideal gas equation of state in ultra-relativistic limit 
(with zero chemical potential so there is no baryon number
conservation equation), $P = \epsilon/3$. 
We use a 3+1  dimensional code using leapfrog algorithm of 2nd 
order accuracy and  QGP ideal gas equation of state for 2 massless 
flavors. The initial conditions here are provided in terms of a 
Wood-Saxon background plus randomly placed Gaussian fluctuations 
of specific widths. We use these initial conditions as it allowed 
for control on the size of initial
fluctuations so that its effects on the locations of acoustic
peaks could be directly studied. We calculate the Fourier coefficients 
for the spatial anisotropies of the energy density (by calculating
net energy contained in a given angular bin) at the initial stage, 
and then using hydrodynamical evolution, calculate the Fourier 
coefficients of the resulting momentum anisotropy in ${\Delta p}/p$
in different angular bins at a later stage. As we mentioned above,
we use a fixed lab frame, and calculate respective power spectra
(of spatial anisotropies and momentum anisotropies, respectively)
using root-mean square values of the respective Fourier coefficients.
For details of the simulations we refer to ref.\cite{hydro}.  

\begin{figure}
\begin{center}
\resizebox{0.7\columnwidth}{!}{%
\includegraphics{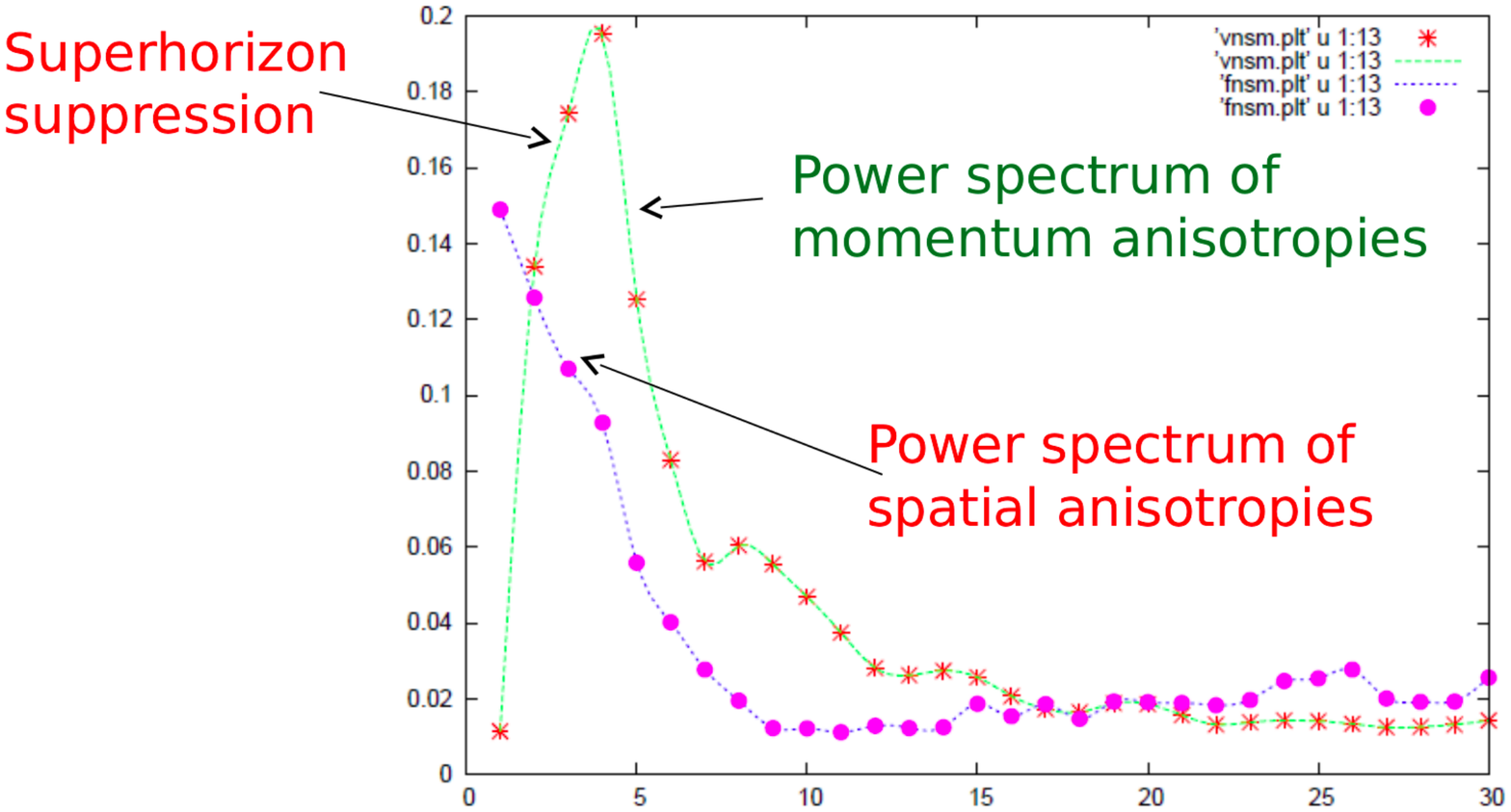} }
\caption{Suppression of low $n$ modes. Note that the power in low $n$ modes
for spatial anisotropies monotonically increases with decreasing $n$ (for
central collision). However, while hydro evolution  converts these shape
anisotropies to momentum anisotropies in a proportional manner for high
values of $n$, it is completely reverse behavior for low values of $n$.
This is exactly what is expected of superhorizon modes for which the hydro
equations do not have enough time to convert the spatial anisotropy to
momentum anisotropy in a proportional manner (not necessarily assuming 
the same proportional factor for all $n$). We also point out hints of
second peak in the power spectrum of  momentum anisotropies
near $n \sim 8-9$. As these results are for short time scale
$\tau = 1.98$ fm, one would not expect full development of
any acoustic peaks.}
\label{fig6}
\end{center}
\end{figure}

 Fig.6 shows plots of these power spectra from the simulation for a 
central collision. Dotted curve with solid dots shows the plot
of initial power spectrum of spatial anisotropies. Dashed curve with
stars shows the power spectrum of resulting momentum anisotropies (at 
proper time $\tau$ = 1.98 fm, we could not evolve the system for large 
times due to certain instabilities for large fluid velocities, see \cite{hydro}
for a discussion). Comparison of the two plots shows important qualitative 
difference for low $n$.  We note that for $n$ larger than about 4, both 
plots show roughly similar pattern. However for smaller $n$, the two 
plots show dramatic difference. Plot for spatial anisotropies  keeps 
rising monotonically with decreasing $n$. However, plot for momentum 
anisotropies (resulting from the spatial anisotropies) shows a drop for
low $n$ values. This suppression of {\it superhorizon modes} was  
predicted in ref.\cite{cmbhic,vnrms} resulting from the fact that these large 
wavelength modes do not get enough time to transfer to momentum anisotropies.
As we discussed above, hints for this suppression for low $n$ values is 
seen in the experimental plots in Fig.4, though latest data shown in
Fig.5 does not seem to support it. There is a hint of a
second peak in Fig.6 in the power spectrum of  momentum anisotropies
near $n \sim 8-9$. As we mentioned above, these results are for short time 
scale, $\tau = 1.98$ fm. One would not expect full development of
any acoustic peaks in such a short time interval. Clearly further efforts 
are needed to probe these very important features. If these are indeed 
found to be present, then suppression of this momentum anisotropy for low 
$n$ values (compared to spatial anisotropies), along with possible existence
of a second acoustic peak, will be very surprising, indicative of a rich 
physics of existence of causal (sound) horizon, suppression of superhorizon 
modes, as well as acoustic oscillations in relativistic heavy-ion collisions.
\section{Initial magnetic field and flow coefficients}
\label{5:1}

 As our focus has been to investigate initial state fluctuations, it is
important to know how to separate them from the effects of hydro evolution. 
What one needs to use is some technique which can, in some way, isolate the 
quantitative values of fluctuations at very early stages. We discuss below 
how to achieve it with the effects of  magnetic field on the power spectrum 
of flow. 

 There has been a tremendous interest in the effects of initial magnetic field 
on the evolution of system in relativistic heavy-ion collisions. In non-central 
collisions, at the center of system, the magnetic field produced by motion of nuclei  
is perpendicular to the {\it reaction plane} (plane formed by the 
impact parameter vector 
and the line of motion of nuclei, $xz$-plane in our case). It is known that one can get 
extremely large magnetic field in these experiments, having strength of the order of 
$10^{14} - 10^{15}$ Tesla (at the center of system), beyond the values anywhere else 
in the observed universe. Some of the main motivations of these studies have been to use 
this magnetic field to directly probe highly non-perturbative physics of QCD such as 
effects of instantons; also the effects of sphaleron at a very high 
temperature and in the out-of-equilibrium state arising from
the so called chiral magnetic effect \cite{cme1}. These processes create 
the domains of gluonic configuration having non-zero integral 
topological charge, typically either +1 or -1. In such domains, a chirality
imbalance is created due to the {\it chiral anomaly} of QCD, i.e. depending 
upon the topological charge of the configuration, either right handed 
chirality dominates or left handed (working in the {\it chiral limit}). 
On the other hand, a strong magnetic 
field aligns the spin magnetic moment of particles along its direction; spin of positive 
charge particles is aligned along the direction of magnetic field and spin of negative 
charge particles in the opposite direction. In the domains of non-zero topological charge, 
depending upon the chirality dominance, this leads the opposite motion of positive and 
negative charge particles, and a local electric current is generated perpendicular to the 
reaction plane. In general there will be many such domains having different 
chirality imbalance, which will therefore have opposite directions of this local current. However since the formation 
of these domains is lead by statistical process, there can be an overall non-zero 
topological charge due to spacetime fluctuations of such topologically non-trivial gluonic 
configurations, which can lead an overall electric charge separation and generate electric 
current perpendicular to the {\it reaction plane}. This is known as the {\it chiral magnetic 
effect} (CME) \cite{cme1}, which is being extensively 
investigated, see review \cite{cme2}. One of 
the main problems in this regard is that the magnetic field is strong only at very early 
times, decaying by few orders of magnitude within a fm time. It was pointed out by 
Tuchin~\cite{tuchin} that since the plasma forms within less than a $fm/c$ time,
the rapidly decreasing magnetic field will induce circular currents in the medium
and as a result the induced magnetic field will survive for much longer times,
the relaxation time depending on the conductivity of the plasma.  But even with the medium 
effects, the extremely large initial values of magnetic field does
not last for any significant time period. Therefore chiral magnetic effect is expected to 
dominantly occur in the pre-equilibrium stage of the collision.   
It has also been argued that the effects of conductivity do not play an 
important role for 
realistic values, and the medium effects are much more suppressed \cite{larry} (see,
also \cite{tuchin3} in this context).

 While this becomes a limitation for studying chiral magnetic effect etc.,
we suggest that this limitation can be used to our benefit in isolating 
the initial distribution of fluctuations from their later evolution. 
Magnetic field, present at the thermalization time, can affect the whole 
evolution of the fluid. It can affect the elliptic flow, and in general affects 
all flow coefficients; the entire power spectrum of flow coefficients can be 
affected by the magnetic field. Interestingly, this is exactly what happens 
for CMBR power spectrum also \cite{cmbrB}. Indeed, that was the motivation for 
some of us to initiate the study of effects of initial magnetic field on flow 
coefficients, in particular on the elliptic flow in relativistic heavy-ion 
collisions, see ref.\cite{v2B}.

We briefly recall the  discussion of effect of magnetic field on elliptic
flow $v_2$ from ref.\cite{v2B}, where it was pointed out that an initial magnetic 
field can enhance elliptic flow; a similar enhancement was also confirmed in an
analysis by Tuchin \cite{tuchinv2}. The basic physics argument in ref.\cite{v2B} is 
as follows. Consider a non-central collision as shown in Fig.1. The moving spectators 
lead to generation of magnetic field $B_0$ along $y$-axis (with impact parameter vector being 
along $x$-axis) in the central region. When a thermalized medium forms, this magnetic 
field remains present, even though with relatively smaller strength, and may get trapped 
inside the plasma. In presence of magnetic field, there are different types of waves 
in the plasma. There are fast magnetosonic waves which are generalized sound waves 
with significant contributions from the magnetic pressure. These waves have speed
$\sqrt{c_s^2+v_A^2}$, where $c_s$=$\sqrt{\frac{\partial p_g}{\partial \epsilon}}$ is the
hydrodynamics sound speed, $v_A$=$\frac{B_0}{\sqrt{4\pi \epsilon}}$ is the 
Alfv$\acute{e}$n speed, $p_g$ is the thermal pressure, and $\epsilon$ is the energy 
density of the plasma. The increment in the speed of such sound waves 
arises, basically because  
under the expansion, distortions of magnetic field lines along the $x$-direction cost energy, 
because of which the equation of state becomes stiffer in this direction, 
causing increment in the sound speed. In the $y$-direction, the sound speed remains 
unchanged, i.e. remains equal to $c_s$. It 
can be seen that with the development of flow from a pressure gradient (using 
Euler's equations \cite{oltr}) the resulting flow velocity is  proportional to sound 
speed square. As the sound velocity becomes larger in the $x$-direction, it follows that 
flow in this direction will be enhanced, while in the $y$-direction it will not change. 
This can lead to the enhancement of the elliptic flow $v_2$.    

 However, the physics of this effect is not that simple, as other factors can be present. 
For example, under certain situations, specially in the high impact parameter regime of 
collisions, 
extent of magnetic field lies beyond the plasma region along the $x$-direction. In that case,
the expansion of a conducting plasma into regions of magnetic field gets hindered. One can 
expect it from Lenz's law: expanding conductor squeezes magnetic flux, which opposes expansion 
of the plasma. Such an argument will imply suppression of $v_2$ due to magnetic field. However, 
as discussed in ref.\cite{v2B}, distortions of magnetic field lines
tries to enhance the flow along $x$-direction.  In general, all such factors
will be present affecting the flow in a complex manner.  As we will see 
below, depending on the situation, specifically, extent of the plasma region
in comparison to the extent of region of strong magnetic field, one of these factors  may 
dominate over the other. Along with these, fluctuations also play important role and 
the final effect is a combination of all these.

The quark-gluon plasma, produced in relativistic heavy-ion collisions, has a finite 
electric conductivity which varies spatially as well as temporally. However, for 
simplicity, we take the ideal magneto-hydrodynamic approximation for this fluid, in 
which the electric conductivity is considered  infinite at each spacetime point. 
The ideal relativistic magneto-hydrodynamics (RMHD) equations are 
\cite{refmhd},\\

\noindent
a) The baryon number conservation equation,
\begin{equation}
\partial_{\alpha}(n u^{\alpha}) = 0.
\end{equation}
b) The energy-momentum conservation equations,
  \begin{equation}
   \partial_{\alpha}\Big( \big(\epsilon +p_g + |b|^2\big) u^{\alpha} u^{\beta}
   - b^{\alpha} b^{\beta} + \big(p_g + \frac{| b |^2}{2}\big)\eta ^{\alpha \beta}\Big) = 0.
  \end{equation}
c) The homogeneous Maxwell's equations,
  \begin{equation}
   \partial_{\alpha}(u^{\alpha} b^{\beta} - u^{\beta} b^{\alpha}) = 0.
  \end{equation}
Here, $n$, $\epsilon$, and $p_g$ are baryon number density, energy density, and thermal 
pressure respectively.  $u^{\alpha}=\gamma(1,\vec{v})$ is the 
four-velocity of the fluid. The Minkowski metric is $\eta_{\alpha \beta}=diag(-1,1,1,1)$,  
the four-vector $b^{\alpha} = \gamma \Big( \vec{v}.\vec{B}, \frac{\vec{B}}{\gamma^2} + \vec{v}(\vec{v}.\vec{B} ) \Big)$, 
and $|b|^2$=$b^{\alpha}b_{\alpha}$. Therefore the total pressure of the fluid is $p =p_g + \frac{| \vec{B}|^2}{2\gamma^2} 
+ \frac{(\vec{v}.\vec{B})^2}{2}$. By following formalism from ref.\cite{refmhd} to solve these equations, we carry out 
ideal relativistic magneto-hydrodynamic simulations for evolution of QGP produced in relativistic heavy-ion collisions. 
We take an initial profile of magnetic field for given impact parameter of the collision at the thermalization time of 
the system, calculated by taking electric field for uniformly charged nuclei, and Lorentz transforming it for their 
opposite motion. We carry out (3+1)-dimensional simulation by using Glauber-like initial energy density for QGP, with 
profile along $z$-direction being Woods-Saxon with appropriate parameters. We show our simulation result of effect of 
magnetic field on the elliptic flow in Fig.7, see ref.\cite{mhd}.

\begin{figure}
\begin{center}
\resizebox{0.5\columnwidth}{!}{%
\includegraphics{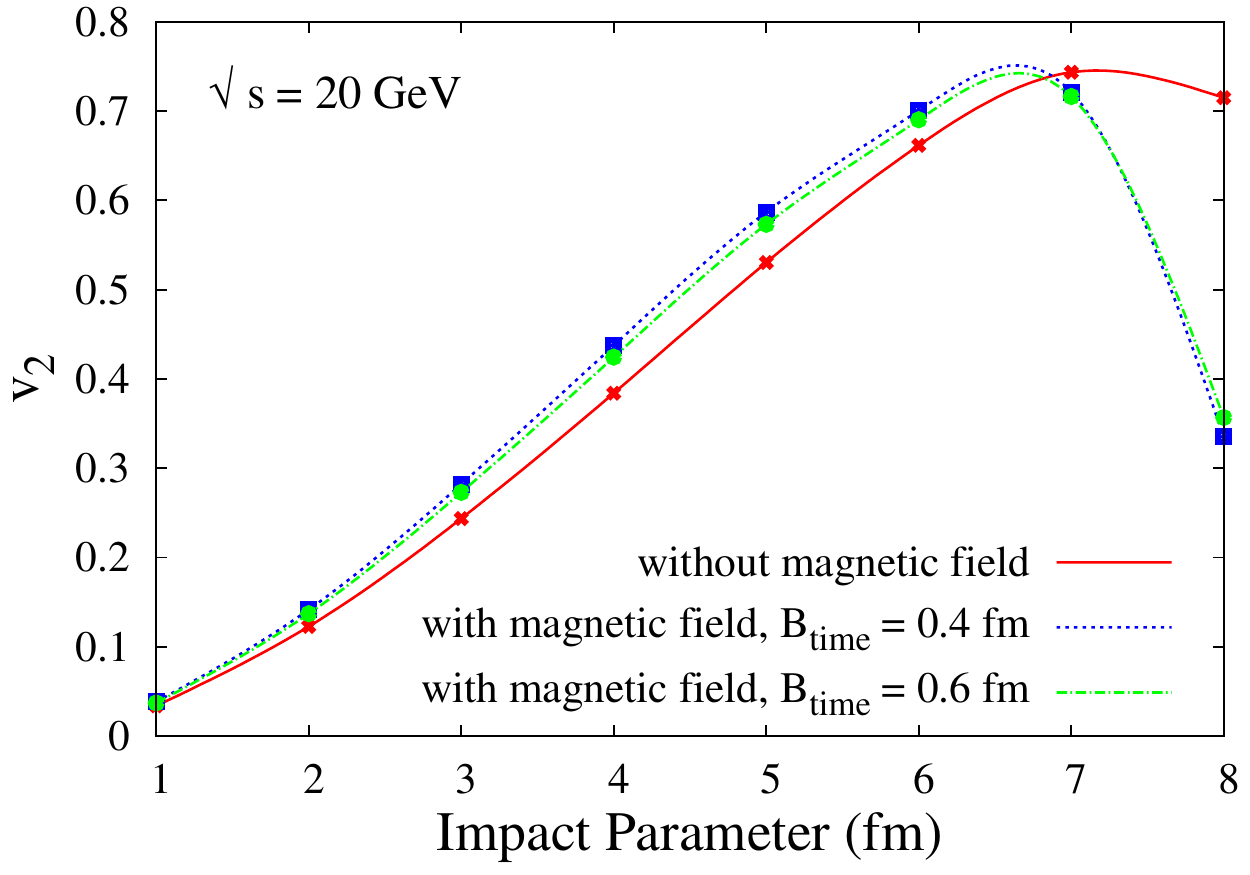} }
\caption{ Effect of initial magnetic field on the elliptic flow \cite{mhd}.}
\label{fig7}
\end{center}
\end{figure}

We see that magnetic field enhances $v_2$ for small impact parameters. However, with increasing impact 
parameter, the enhancement increases first and then decreases. Eventually, at very large impact 
parameters, magnetic field suppresses the elliptic flow. This non-trivial effect of magnetic field on 
$v_2$  arises due to the following reasons. If magnetic field is almost entirely contained within the 
plasma region, elliptic flow gets enhanced by the magnetic field. This is only possible for small values of 
the impact parameter. This is in accordance with the argument of having a stiffer equation of state 
along $x$-direction due to the magnetic field along $y$-direction. However, if the magnetic field 
extends well beyond the plasma region along $x$-direction, then elliptic flow is suppressed 
by the magnetic field due to the Lenz's law. This situation arises when impact parameter 
is large. 

Refs.\cite{Inghirami11,Inghirami12} also study the effects of magnetic field on elliptic 
flow by performing ideal RMHD simulations. In \cite{Inghirami12} it is shown
that a strong 
magnetic field can enhance elliptic flow, where, the magnetic field generated by electric current 
arising due to CME ($I_{cme}$) in the pre-equilibrium stage also has been considered along with the classical 
origin of magnetic field. The classical origin of magnetic field is 
calculated in a medium with a non-zero electric conductivity.
The total magnetic field profile arising from these two sources is set as the initial condition for the 
evolution of the fluid. The parameter which enters for the calculation of magnetic field of CME origin is the 
{\it chiral magnetic conductivity} $\sigma_{\chi}$ \cite{cmc}, which is a proportionality constant of 
$I_{cme}\propto \vec{B}$. It is found that the magnetic field generated due to the CME has opposite effects on 
the elliptic flow, i.e. it has tendency to suppress the elliptic flow even in low impact parameter regime 
\cite{Inghirami12}. In that work, the dependence of electric conductivity and $\sigma_{\chi}$ on the initial 
magnetic field profile and its effect on elliptic flow has also been studied. 
In ref.\cite{royMHD}
by performing reduced-magnetohydrodynamical simulations for expansion of hot and dense nuclear matter in (2+1)-dimensions,
the enhancement of $v_2$ is reported. In ref.\cite{bv2ref2} also, effect of an inhomogeneous magnetic field on 
the transverse flow has been investigated. 

In our simulation \cite{mhd}, we also find that fluctuations in the initial energy density can lead to
temporary increase of magnetic field in some fluid regions due to flux-rearrangement by evolving initial 
state density fluctuations, which can push flux lines, leading to temporary and localized concentration 
of flux lines. This will be important for CME which is sensitive to locally strong magnetic field (instanton 
size regions). 

 We now show an important qualitative effect of magnetic field on
the power spectrum of flow coefficients. Fig.8 shows the power
spectrum of flow for magnetic field with strength $5 m_\pi^2$ \cite{mhd}.
As this plot is for a strong magnetic field, simulation
could be carried out only for  short time of 0.6 fm.
We see a pattern of different powers in even and odd $v_n^{rms}$ coefficients at low $n$. 
This is expected from the reflection symmetry about the magnetic field direction 
if initial state fluctuations are not dominant. This is a qualitatively distinct 
result with unambiguous signal for the presence of strong magnetic field during 
early stages. 
\begin{figure}
\begin{center}
\resizebox{0.5\columnwidth}{!}{%
\includegraphics{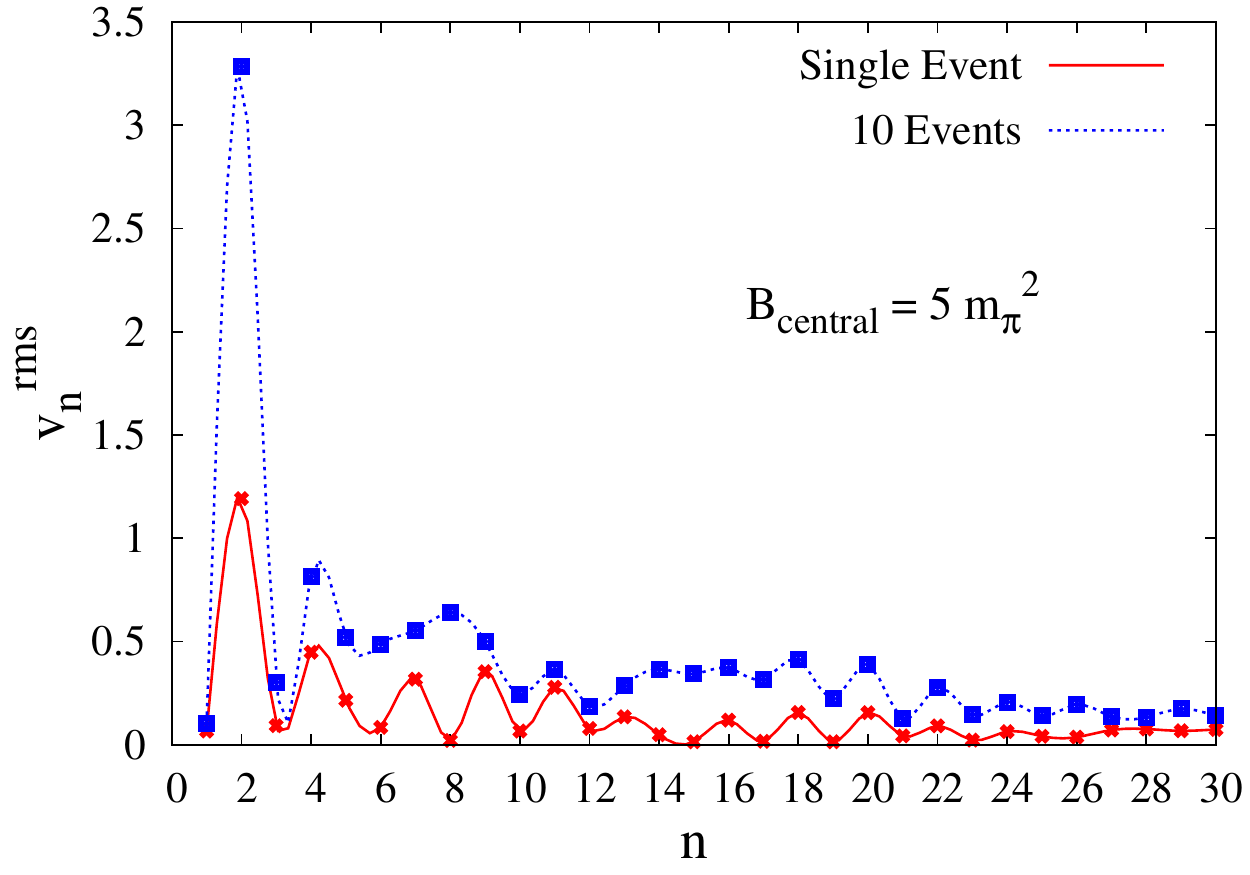} }
\caption{Plot of $v_n^{rms}$ for magnetic field with
strength $5 m_\pi^2$. Even-odd power difference is seen in first
few flow coefficients as fluctuations wash out the effect for
large $v_n$s \cite{mhd}. There is no hint of such features in
the data in Figs. 4,5 which could be due to strong initial
fluctuations, or simply that strong enough magnetic field
does not survive for any significant time.}
\label{fig8}
\end{center}
\end{figure}

Note that the even-odd pattern is seen in Fig.8 for only first few flow 
coefficients as fluctuation effects wash out the effect for larger $v_n^{rms}$ 
for the event average over 10 events. As fluctuations are necessarily
present at the initial stage itself, this signal will be in general
suppressed in the final flow power spectrum. This could be one possible 
reasons why there is no hint of such features in the data in Fig.4 and
Fig.5. Of course, it is also possible that strong enough magnetic field 
may not last for significant time for this feature to develop
sufficiently. To illustrate the effect of initial fluctuations on this
feature, we show in Fig.9 flow fluctuations for a smooth isotropic plasma
region (without any fluctuations) in the presence of magnetic field.
We now take a more reasonable value of magnetic field strength
equal to $m_\pi^2$. Due to smaller magnetic field and smooth
plasma profile, the evolution could be run up to 3 fm time (after which
boundary effects could not be neglected). We see a strong even-odd
power difference in the power spectrum even for large $n$ values.
\begin{figure}
\begin{center}
\resizebox{0.5\columnwidth}{!}{%
\includegraphics{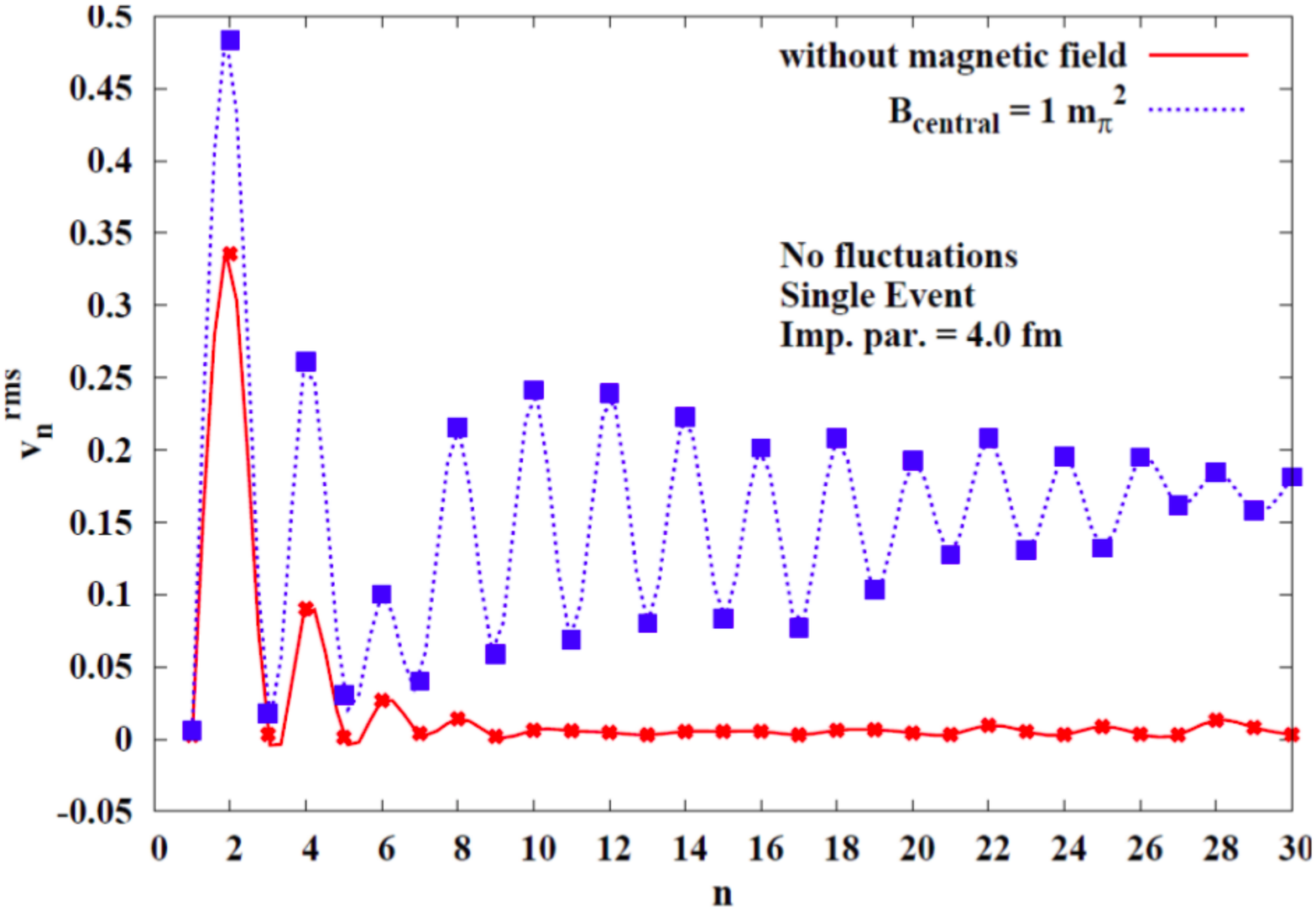} }
\caption{ Plot of $v_n^{rms}$ for magnetic field with
strength $m_\pi^2$. Here we consider isotropic region with smooth
plasma profile without any fluctuations.  Strong difference in
the power of even and odd values of $v_n^{rms}$ are present
arising from the effect of magnetic field \cite{mhd}.}
\label{fig9}
\end{center}
\end{figure}

The suppression of this qualitative even-odd signal for flow power spectrum 
\tct{(e.g. absence of such a feature in Figs.4,5)} provides us an independent 
probe of initial state 
fluctuations. As the magnetic field is strong only for very early stages,
the evolution of flow power spectrum during those stages will be a result
of complex interplay of magnetic field effect, producing even-odd power 
differences, and existence of fluctuations, which tend to suppress these 
qualitative features. A comparison with detailed
simulations should be able to shed some light on the nature of fluctuations
during very early stages when the magnetic field was strong.

Now we discuss very briefly another aspect of QCD matter which may affect $v_2$. An ideal MHD fluid has a property 
of diamagnetism, which opposes any change in the strength of magnetic field if fluid is at rest; magnetic flux lines 
are conserved in this fluid. As we mentioned earlier, in this fluid, an additional momentum anisotropy arises due to 
larger sound speed along $x$-direction. In contrary to this, it is shown 
in ref.\cite{paraSqeez}, that the QCD 
medium has a property of paramagnetism, which supports changes in the strength of magnetic field. It is then argued 
that this feature may create an additional spatial anisotropy in the fluid simply because such fluid will move towards 
the region of stronger magnetic field, which may make plasma more squeezed along $x$-direction \cite{paraSqeez}. This 
process is named as {\it paramagnetic squeezing}. This additional spatial anisotropy may affect $v_2$ depending upon the 
impact parameter of the collisions \cite{paraSqeez}. Note that in ref.\cite{bv2ref1}, suppression in $v_2$ due to 
this effect has been reported. However in that work a magnetic field 
profile with non-zero divergence was used, so results may not be conclusive.

\subsection{Flow anisotropies and superfluid phases of QCD}
\label{5:2}

 It turns out that this qualitative behavior of even-odd power difference
 for flow coefficients can also arise from an entirely different source.
 If there are superfluid vortices present during early stages of low energy heavy-ion 
 collisions, they can also lead to such features \cite{qcdsf}. Of course, in that 
 case there will be additional signals, such as a very strong elliptic flow even in the central
 collisions, negative elliptic flow for some specific configuration of vortices in non-central 
 collisions etc.  \cite{qcdsf}, which can be used to differentiate from the effect of
 magnetic field sourced even-odd power difference. There is a remarkable
 variety of exotic phases of QCD at very large baryon density, e.g.
 color flavor locked (CFL) phase, 2SC phase, crystalline superconductivity
 etc. These are color superconducting phases of QCD arising
 from di-quark condensates, with quarks near the Fermi surface forming
 Cooper pairs at very high baryon density \cite{cfl}. Some of these phases,
 e.g. CFL phase, lead to superfluidity. Interestingly, even at relatively
 low baryon densities the nucleonic superfluidity with neutrons forming 
 Cooper pairs (for protons one gets superconductivity) also exists, which is
 typically found in the interiors of neutron stars.

  Such superfluid phases may become accessible in relatively low energy
  heavy-ion collisions, e.g. at FAIR and NICA, and possibly at the beam energy
  scan program of RHIC. Any transition to superfluid phase will invariably
  lead to formation of superfluid vortices whose initial number density can
  be estimated from reasonably model independent topological arguments (see,
  \cite{qcdsf}). It is clear that any superfluid vortex at the initial stage
  will dramatically affect the resulting flow pattern. This was investigated in
  ref.\cite{qcdsf} using relativistic hydro simulations, incorporating initial
  vortex configurations and several qualitatively new features were found.
  For example, a strong even-odd power difference in the power spectrum
  of flow coefficients was found, similar to shown in 
  Fig.8,9. Along with that strong elliptic flow in 
  central collisions, and negative elliptic flow in non-central 
  collisions were also found, where different
  possibilities arise for different initial vortex configurations.
  Thus, with these, one can distinguish the source of any even-odd power
  difference from the effect of initial magnetic field. More importantly, the
  two effects arise in entirely different regimes of QCD phase diagram.
  Strong magnetic field only occurs at ultra-relativistic collisions, e.g.
  at highest energies of LHC which invariably has very small baryonic
  chemical potential associated with the produced QGP. So there is no
  possibility of any superfluid QCD phases arising in that energy regime. On the
  other hand, low energy collisions at FAIR, NICA, BES program of RHIC, which
  may have a high value of baryon chemical potential are not
  expected to have very high magnetic fields which could lead to any
  significant even-odd effect for the flow power spectrum.

  \section{Conclusions and future directions}
  \label{6:1}
   We have provided a short review of a 
   very specific topic, focusing on the power spectrum
    of flow fluctuations in relativistic heavy-ion collisions. The thermalized
    medium formed from collision of two heavy nuclei is viewed exactly in
    the same manner as the initial matter-energy density in the universe with
    associated density fluctuations. These density fluctuations get imprinted
    into the final particle momentum distributions, just as for the universe
    the initial primordial density fluctuations manifest in final photon
    distributions leading to CMBR power spectrum. With that lesson in mind from
    the universe, the power spectrum of flow fluctuations becomes an excellent
    probe for the initial state fluctuations in heavy-ion collisions.  
    The physics
    underlying the evolution of initial density fluctuations is very similar in
    both cases, simply governed by relativistic hydrodynamical equations in
    expanding plasma (though expansions are different in both cases). The only
    important difference between the two cases being absence of gravity for
    RHICE. However, it is easy to see that the presence of acoustic peaks is
    independent of the presence of gravity, simply resulting from sound
    modes in a plasma and the superhorizon density fluctuations
    are necessarily present in RHICE at the initial stage. One of the most
    important features expected in the power spectrum of flow coefficients
    is the suppression of long wavelength modes or the low $n$ flow
    coefficients. There seems clear experimental evidence for the suppression
    of long wavelength fluctuations (lower $n$ 
    flow coefficients) in experimental
    data and it is important to focus on these to probe long range correlations
    at initial stage. This will shed light on the presence of long scale
    correlations in initial parton distributions (which will be probed by the
    upcoming electron-ion collider), and also on the size of sound horizon
    at the freezeout stage (just like for CMBR, the first peak signals the size
    of causal horizon at the surface of last scattering). We have also discussed
    how the existence of strong magnetic field in very early stages of plasma
    evolution (which rapidly decays, even with medium effects) can be used to
    isolate the initial values of density fluctuations from the effects of
    their subsequent evolution. This is in terms of a qualitative effect of
    strong magnetic field leading to difference in power of even-odd flow
    coefficients. As initial fluctuations suppress these effects, therefore 
    with proper numerical simulations one may be able to use the suppression 
    of these qualitative feature to provide us an independent probe of initial 
    state fluctuations.

\section*{Acknowledgement}
    We acknowledge useful discussions with Sanatan Digal,
    Minati Biswal, and Abhishek Atreya.

\end{document}